\documentstyle[preprint,prb,epsfig,aps]{revtex}
\tightenlines
\begin{document}
\draft
\title{Synthesis and Processing of $MgB_2$ powders and wires}
\author{C. E. Cunningham,\thanks {On leave from Dept. of Physics, Grinnell College, 
Grinnell, IA 50112} C. Petrovic,  G. Lapertot,\thanks {On leave form Commissariat a 
l'Energie Atomique DRFMC-SPSMS, 38054 Grenoble, France} S. L. Bud'ko, F. Laabs, W. Straszheim, D. K. Finnemore, and P. C. Canfield}
\address{Ames Laboratory, U.S. Department of Energy and Department of Physics and Astronomy\\
Iowa State University, Ames, Iowa 50011}
\date{\today}
\maketitle
\begin{abstract}

Sintered powders and wires of superconducting $MgB_2$ have been fabricated under a variety 
of conditions in order to determine details of the diffusion of the $Mg$ into $B$ and to study the types of defects that arise during growth.  For samples prepared by exposure of 
boron to $Mg$ vapor 
at $950^{\circ }C$, the conversion of particles of less than $100~\mu m$ size particles to 
$MgB_2$ is complete in about $2~h$. The lattice parameters of the $MgB_2$ phase 
determined from X-ray are independent of the starting stoichiometry and the time of reaction.  
Wire segments of $MgB_2$ with very little porosity have been produced by reacting 
$141~\mu m$ diameter boron fibers in an atmosphere of excess $Mg$ vapor at $950^{\circ }C$.
Defects in the reacted fibers are predominantly the voids left as the boron is converted 
to $MgB_2$.
\end{abstract}


\section{Introduction}
The discovery of superconductivity in the intermetallic compound, $MgB_2$ near  
$40~K$\cite {1} has prompted both fundamental studies of the mechanism causing 
superconductivity\cite {2} and practical studies of supercurrent transport at grain 
boundaries.\cite {3,4}  Bud'ko 
and co-workers\cite{2} have shown the existence of a boron isotope effect in that 
$T_c~\sim ~M^{\alpha }$ with $\alpha =0.26$, where $T_c$ is the superconducting 
transition temperature and $M$ is the isotopic mass.  This result implies that the 
electron-phonon interaction is important in the determination of $T_c$ and that 
light elements with high phonon frequencies might be important in the search for 
new metallic superconductors with high transition temperatures.  Energy gap measurements\cite {5} 
and neutron scattering measurements of the phonon spectrum\cite {6} all point to a conventional 
mechanism for the superconductivity.  On the practical side, Canfield and co-workers\cite {7} 
have shown a method to convert commercially available boron fibers\cite {8} into very low 
resistivity $MgB_2$ wire and Takano and co-workers\cite {9} have measured the current 
carrying capacity of hot-pressed pellets of the material.  Other measurements of flux pinning and 
intergrain coupling have been made using studies of flux creep.\cite {10}

In this paper, we report a systematic study of the vapor diffusion method of sample preparation 
for both microcrystalline powders and boron ($B$) fibers, which are available 
commercially in kilometer lengths.  The primary variable in the study is time, and for the wire 
samples the effect of defects in the starting $B$ fibers.  Of particular interest is the role of the 
tungsten-boride core in creating cracks and transmitting $Mg$ along the length of the fiber.  

\section{Experiment}

For powder synthesis, we place a stoichiometric mixture of 
bright lumps of $Mg$ ($99.9\% $) and fine 
crystalline $^{11}B$ powder ($99.5\%$ ) that is less than $100~\mu m$ in diameter 
into a $Ta$ tube whose ends are sealed under a partial atmosphere 
of $Ar$.  For the wire synthesis, we cut $B$ fiber lengths several $cm$ long and either 
$100~\mu m$ or $141~\mu m$ in diameter and place them in $Ta$ tubes under 
a partial pressure of $Ar$ with excess 
$Mg$.  These starting $B$ fibers have a tungsten-boride core about $15~\mu m$ in diameter 
with breaks every few $cm$.\cite {8}   

After the $Ta$ tubes are sealed, they are in turn sealed in quartz ampoules 
with $\sim 175~mbar$ of $Ar$ and placed
in a box furnace preheated to $950^{\circ }C$.
It is believed that using a preheated furnace inhibits the growth of higher 
boron phases.   After reaction, the samples are rapidly cooled by putting the quartz tube 
in running cold water.  The formation of $MgB_2$ from $B$ causes the powders to expand and bow the $Ta$ tube outward\cite {2} and causes $100~\mu m$ diameter wires to expand by 
more than $30\%$ in diameter.\cite {7}  

Samples are characterized by X-ray diffraction, magnetization, and electrical resistivity.  
X-ray diffraction spectra are taken with $Cu~K_{\alpha }$ radiation in a Scintag 
diffractometer where the scan
angle $2\theta $ varies from $20^\circ $ to $90^{\circ }$with a
scan step of $0.020^{\circ }$.  Magnetization is measured in a Quantum Design 
$MPMS-5$ magnetometer with a $6~cm$ sample travel in a 
magnetic field of $25~Oe$ in a zero-field-cooled mode.
Resistivity is measured with a
standard four probe technique using an $LR-700$ resistance bridge.

\section{Results and discussion}

The lattice constants of the $MgB_2$ phase, 
determined from X-ray data, change relatively little with  time of reaction for  
powders and have values close to those  
previously reported.\cite {11}  The dynamics of the growth of the $MgB_2$ phase are  
illustrated by the study of powder diffraction X-ray 
data after different times of reaction in Fig. 1.  The major $B$ peak at $38^{\circ }$, marked by the solid line on the figure, is small and goes away during processing. 
 For particle sizes used here (less than $100~\mu m$), the 
$MgB_2$ peaks grow quickly between $30~min$ and $60~min$ and remain relatively 
unchanged from $120~min$ to $240~min$.  The $Si$ peaks used for a standard have been removed from these spectra.  It should be noted that when the reaction ampoule was placed in a 
furnace, heated to $950^{\circ }C$, left for $120~min$, and quenched, the powder X-ray 
spectra showed many second phase peaks.   Zero-field-cooled magnetization measurements for these same 
samples show the transition width narrowing from $15~min$ to $30~min$ and then 
remaining relatively constant at about $0.5~K$.

Electrical resistivity of the sintered pellets is compared to the wires in Fig.2.  The inset in Fig. 2 
shows an expanded view near the transition temperature and indicates values at $40~K$ 
of about $1.5~\mu ohm-cm$ for the pellets and $0.4~\mu ohm-cm$ for the wire segments.  These 
processing methods give very low electrical resistivity in the normal state for both pellets and wires.    Because both the relatively open structure of  the pellet and the dense structure of the wire samples have comparable residual resistivity ratios, $RRR\geq 20$, we think the lower resistivity values for the wire sample are due to its higher density.

In order to study the synthesis of $MgB_2$ wire, a series of exposure time, t,  of the boron 
fibers in $Mg$ vapor was measured for $15~min\leq t\leq 240~min$.
For the processing of the $100~\mu m$ diameter fibers, the scanning electron micrographs 
in Fig. 3a and 3b illustrate the behavior seen.  In these micrographs, $Mg$ has a much higher atomic number than $B$ so the shading 
is a good indication of $Mg$ content.   In Fig. 3a, a length of  fiber reacted for $60~min$ has been 
polished through a section that misses the tungsten-boride core in the middle of the fiber.  
The $Mg$ concentration is very high at the outside surface 
region marked (1), and there is a clear front of $Mg$ moving across 
the fiber.  At the boundary between the regions marked (2) and (3), energy dispersive X-ray 
spectra (EDS) show that the $Mg$ level abruptly falls by about $40\%$.  Region (3)
has a fairly high 
$Mg$ content, even though it has not transformed to the $MgB_2$ phase.  As 
shown by the diameter lines, the diameter of this section has grown 
by more than $10\%$ from the original 
$100~\mu m$.

A different section of fiber reacted for $60~min$ in the same run as in Fig. 3a  is shown in Fig. 3b.  
Here the polishing just barely exposes the tungsten-boride core along the centerline.  This 
SEM micrograph shows many of the common defects.  The dark line along the center is a void 
region, which often occurs near the tungsten-boride core, and an EDS spectrum shows 
very strong $W$ lines in this region.  The bright region marked (A) is an $MgB_2$
region and the dark region marked (B) is an unreacted $B$ region.  If optical pictures are taken, 
the $B$ regions stand up above the $MgB_2$ regions because they are harder and polish less 
rapidly.  Figure 3c shows an EDS line scan measurement of the $Mg$ concentration 
taken along the dark 
line running from upper left to lower right in the left-center portion of Fig. 3b.  
The $Mg$ count on Fig. 3c drops from about $1.4$ to about 
$0.9$ going from the bright region to the dark region.  The tick mark on the line in Fig. 3b corresponds to the vertical line in Fig. 3c and 
to a point where the scan moves from a dark region back into a bright region near 
the core.  Magnesium seems to diffuse very rapidly along the core region and diffuse outward 
from there to form 
$MgB_2$.  The void region shown by the dark area in the upper left corner of Fig. 3b is unusually large, and the bottom of the void is clearly seen in the SEM.  Voids are commonly about $10-20~\mu m$ across, and the 
bottom of the void generally can be seen in the SEM.  
The diagonal regions of high $Mg$ content we believe are 
cracks where the $Mg$ can diffuse rapidly and start the formation of $MgB_2$.

A $B$ fiber reacted for $30~min$ is shown in Fig. 4a.  Again the polishing depth does not cut 
through the core.  The sample has much less conversion to $MgB_2$, shown by the bright spots.  
EDS measurements show the highest $Mg$ content in the bright spot marked (C), substantially 
less $Mg$ near the edge at the point marked (A), and even less at the point near the center 
marked (B).  

Another $B$ fiber reacted for $15~min$ is shown in Fig. 4b.  In the left edge of 
this fiber, the tungsten core shows through, and the diameter of the fiber is still close 
to $100~\mu m$.  The EDS data show only a small amount of $Mg$  near the edge of the fiber 
and practically none in the center of the fiber.

When exposure time, $t=120~min$, the $100\mu m$ starting diameter fibers are fully reacted and 
end up having a final diameter of about $160~\mu m$.  To examine how the process scales with 
initial fiber size, a $141\mu m$ diameter fiber was exposed for a much longer time, $t=36~h$, 
with the results shown in Fig. 5.  
The conversion to $MgB_2$ has increased the diameter of the wire to greater than 
$180\mu m$, as shown by the diameter measurements.  There are no second phase particles 
large enough to be seen in the SEM micrograph, and the fiber is relatively void free.

\section{Conclusions}

Low resistivity, high purity powders and highly dense wires of $MgB_2$ can be 
synthesized by exposure of boron powder and boron fibers to $Mg$ vapor at $950^{\circ }C$ 
for adequate exposure time.
Comparable results are seen for both $100~\mu m$ and $141~\mu m$ 
diameter fibers.   For $B$ fibers that are placed in a furnace preheated to $950^{\circ }C$, 
the $MgB_2$ phase begins to form quickly after about $15~min$ and is about $50$ percent 
complete after an hour.  Diffusion of $Mg$ along the tungsten-boride core and along voids 
or cracks in the fiber are important means of transport.  As the $MgB_2$ phase forms, the diameter of the fiber grows from about $141~\mu m$ to about $190~\mu m$.  As reported 
earlier,\cite {7} the wire segments provide a highly conducting material with a normal state 
resistivity of about $0.4~\mu ohm-cm$ and a superconducting critical current density of 
about $20,000~A/cm^2$ at $1~T$ and $20~K$.

\section{Acknowledgments}

Work is supported by the U.S. Department of Energy, Basic Energy Sciences, 
Office of Science, through the Ames Laboratory under Contract No. W-7405-Eng-82.

\vfil\eject
\begin{figure}
\epsfxsize=0.9\hsize
\vbox{
\centerline{
\epsffile{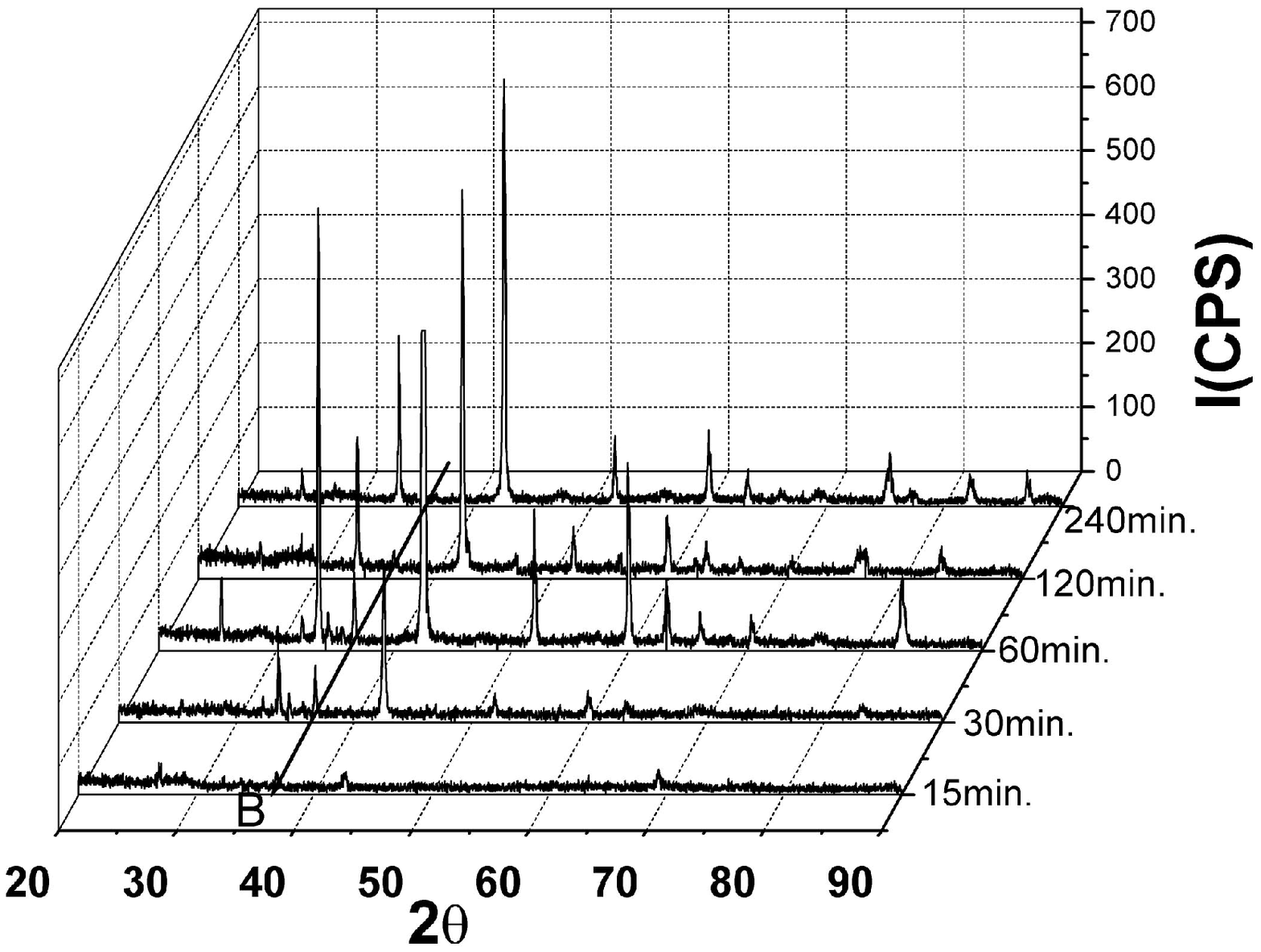}
}}
\caption{X-ray spectra for boron powder reacted for different times.} 
\label{F1}
\end{figure}
\vfil\eject
\begin{figure}
\epsfxsize=0.9\hsize
\vbox{
\centerline{
\epsffile{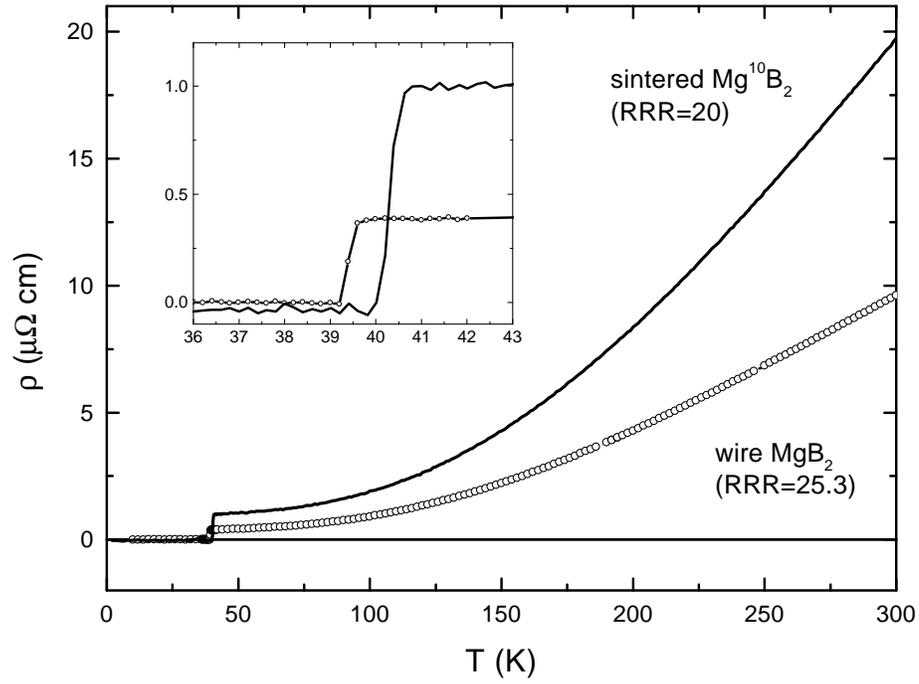}
}}
\caption{Comparison of electrical resistivity of a sintered pellet and a reacted wire segment.}
\label{F2}
\end{figure}
\vfil\eject
\begin{figure}
\epsfxsize=0.9\hsize
\vbox{
\centerline{
\epsffile{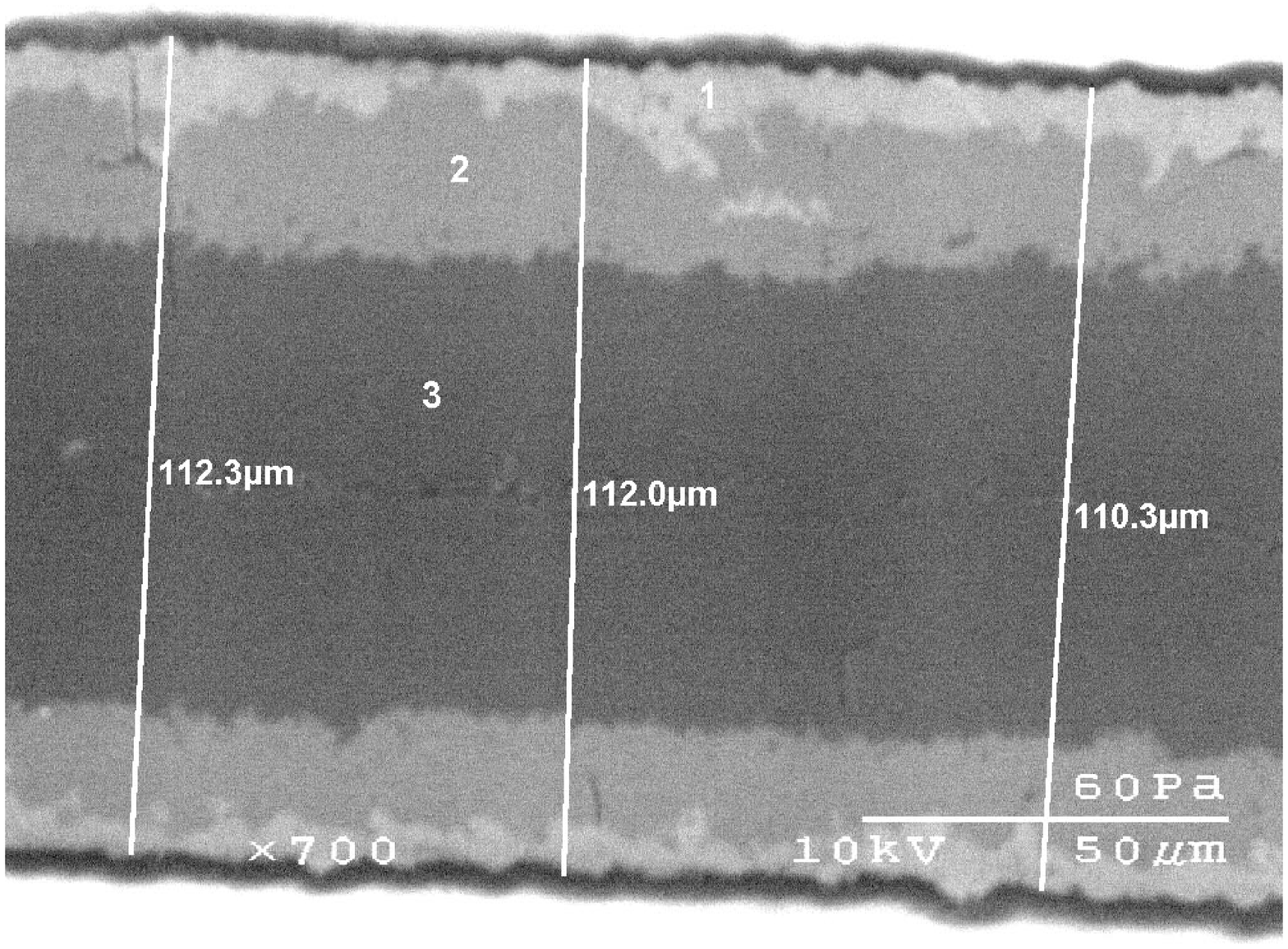}
}}
\vfil\eject
\epsfxsize=0.9\hsize
\vbox{
\centerline{
\epsffile{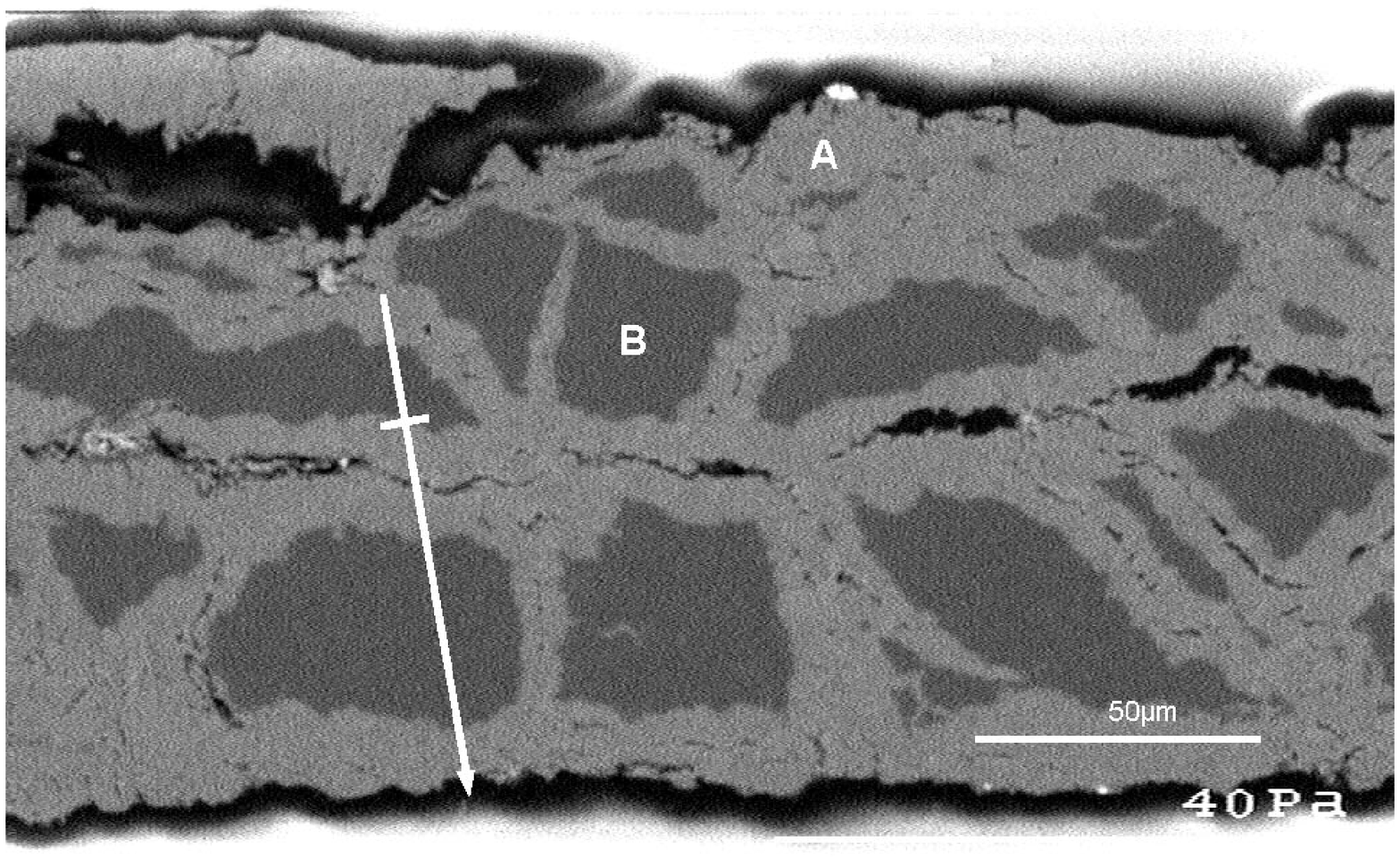}
}}
\vfil\eject
\epsfxsize=0.9\hsize
\vbox{
\centerline{
\epsffile{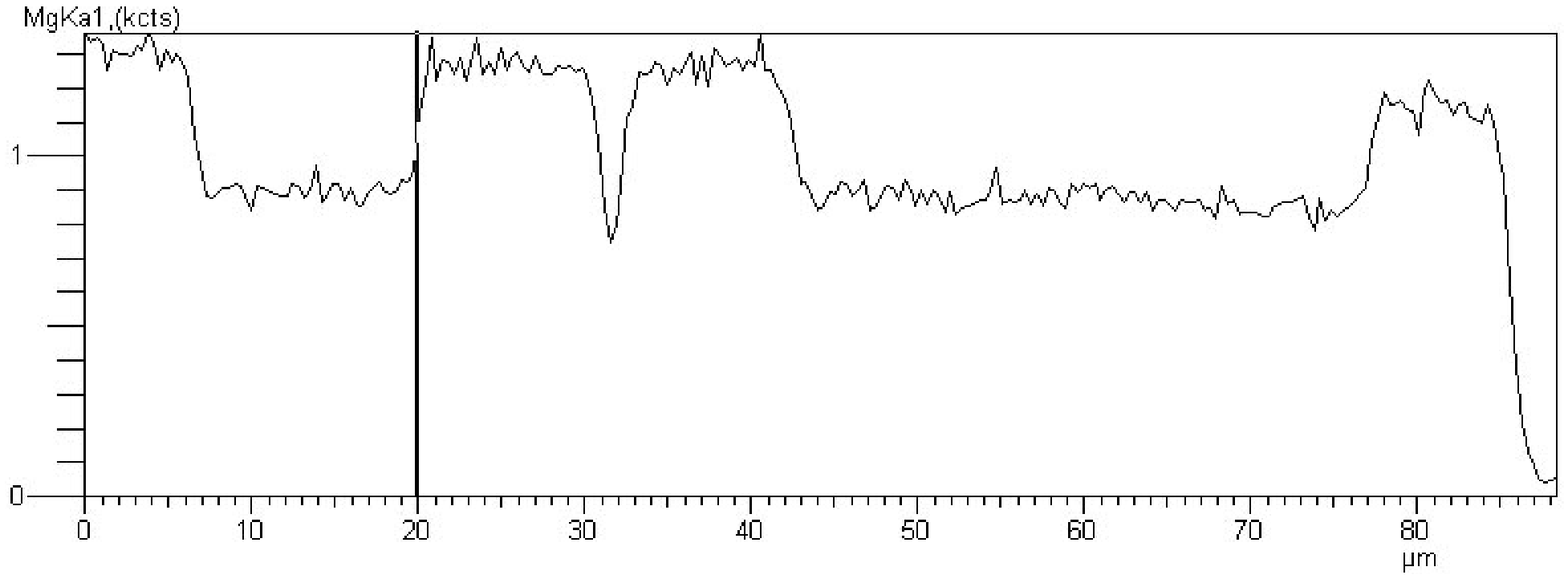}
}}
\caption{SEM micrograph of two segments of $100~\mu m$ diameter boron fibers after 
$60~min$ in $Mg$ vapor at $950^{\circ }C$.  Fig. 3c is an EDS scan along the line 
drawn in Fig. 3b. }
\label{F3}
\end{figure}
\vfil\eject
\begin{figure}
\epsfxsize=0.9\hsize
\vbox{
\centerline{
\epsffile{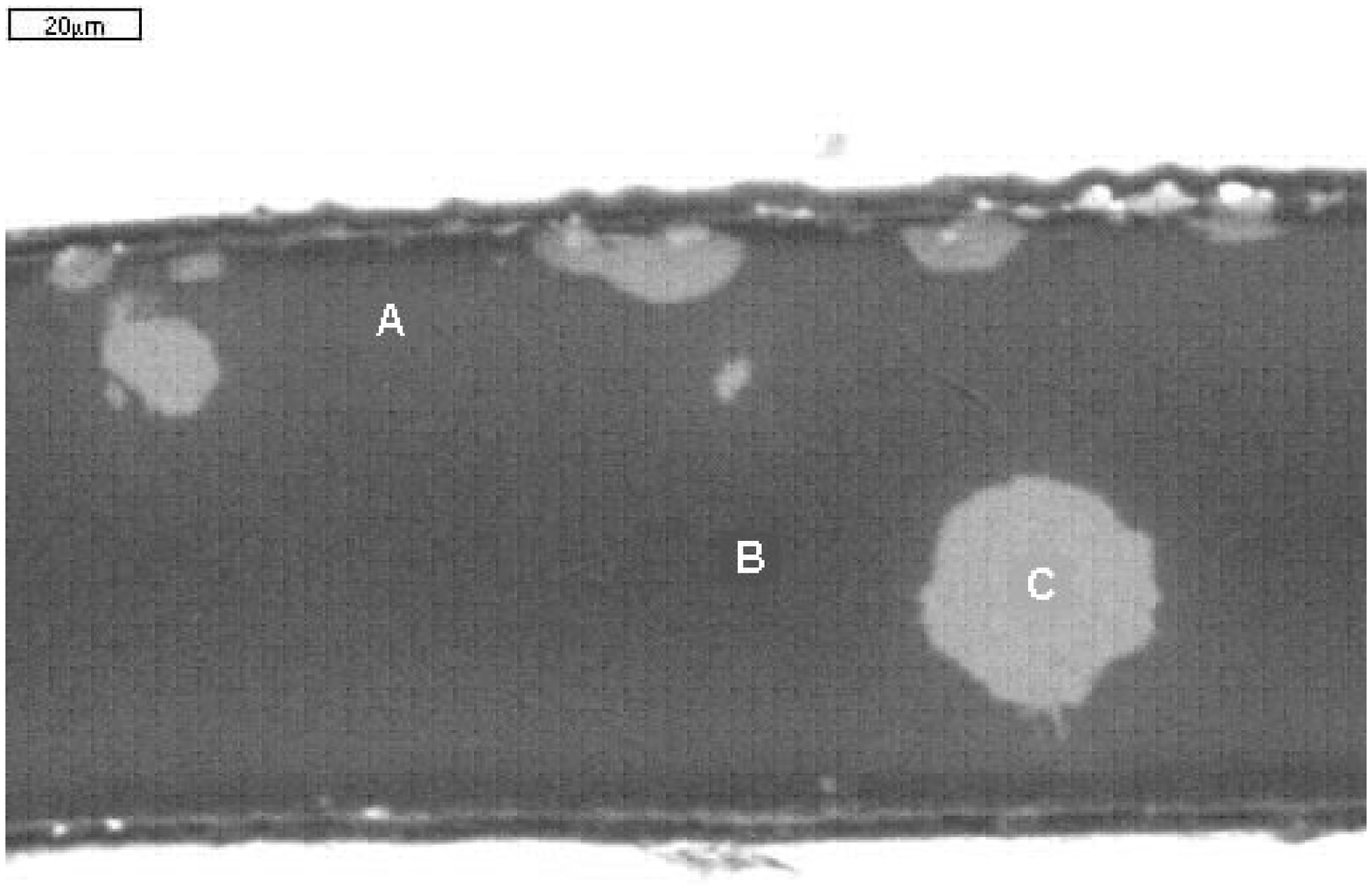}
}}
\vfil\eject
\epsfxsize=0.9\hsize
\vbox{
\centerline{
\epsffile{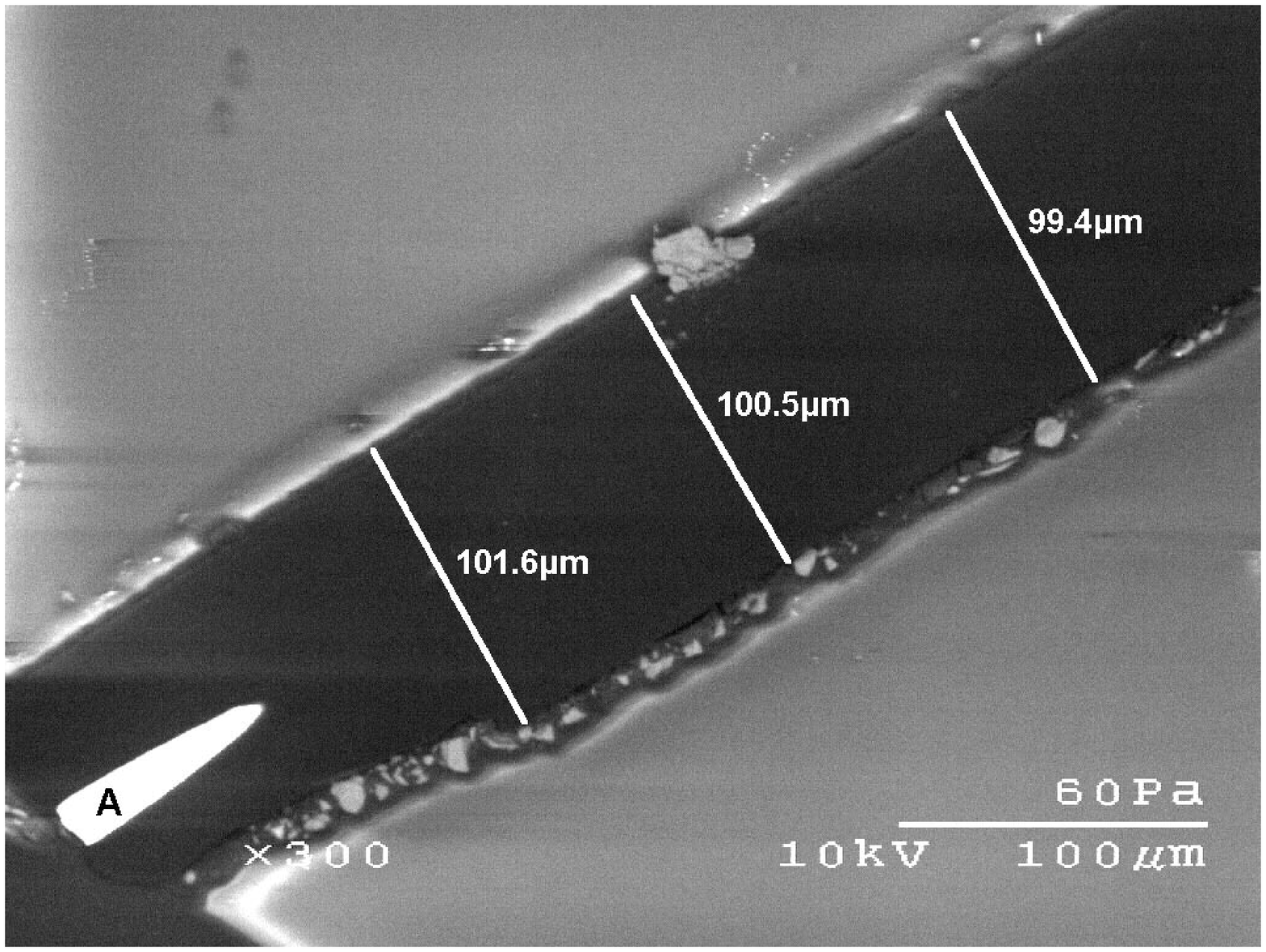}
}}
\caption{SEM micrographs of $100~\mu m$ boron fibers reacted
for a) $30~min$ and
b) $15~min$.}
\label{F4}
\end{figure}
\vfil\eject
\begin{figure}
\epsfxsize=0.9\hsize
\vbox{
\centerline{
\epsffile{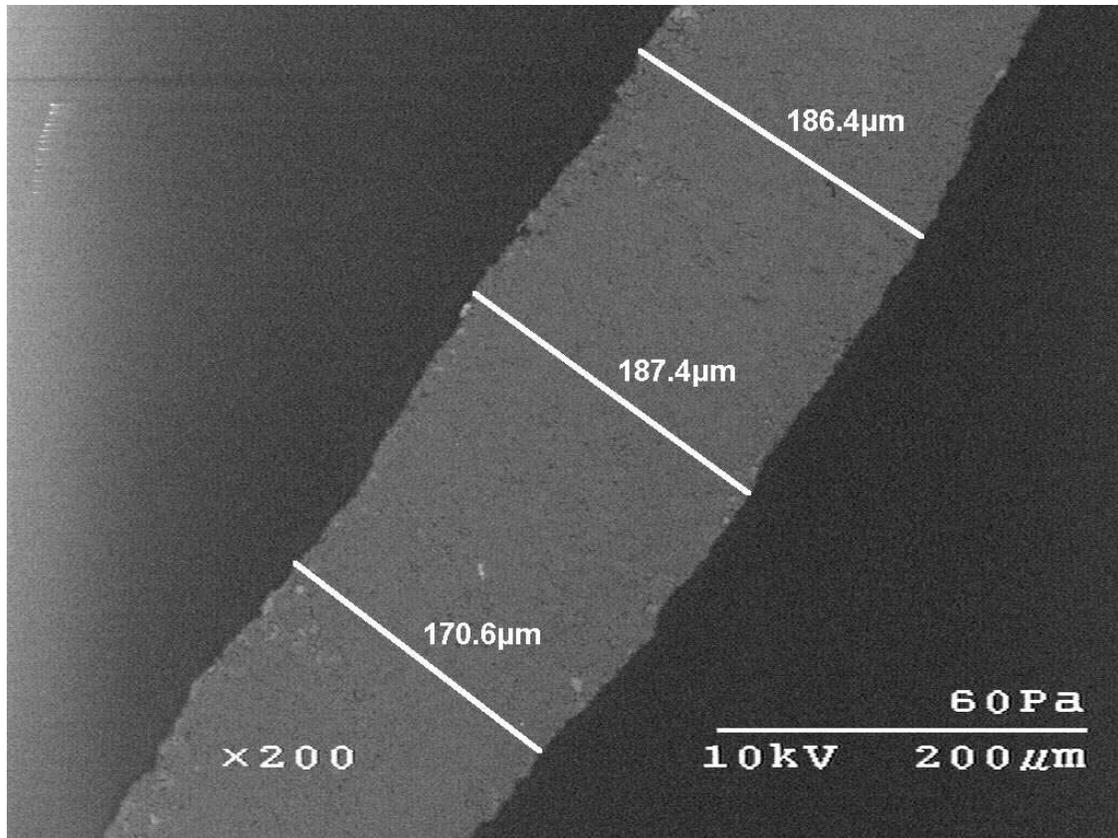}
}}
\caption{SEM micrographs of a
wire segment for the larger diameter boron fiber that is
fully reacted.}
\label{F5}
\end{figure}

\vfil\eject
\end{document}